\documentclass[a4paper,twoside]{article}

\usepackage{epsfig}

\usepackage{subfig}
\usepackage{calc}
\usepackage{amssymb}
\usepackage{amstext}
\usepackage{amsmath}
\usepackage{amsthm}
\usepackage{multicol}
\usepackage{pslatex}
\usepackage{apalike}
\usepackage{SCITEPRESS}
\usepackage[font={small}]{caption}
\usepackage{longtable,booktabs}

\usepackage{balance}

\usepackage{ifxetex,ifluatex}
\ifxetex
  \usepackage[setpagesize=false, 
              unicode=false, 
              xetex]{hyperref}
\else
  \usepackage[unicode=true]{hyperref}
\fi
\hypersetup{breaklinks=true,
            bookmarks=true,
            pdfauthor={},
            pdftitle={Comparison of an open-hardware electroencephalography amplifier with medical grade device in brain-computer interface applications},
            colorlinks=true,
            citecolor=blue,
            urlcolor=blue,
            linkcolor=magenta,
            pdfborder={0 0 0}}
\usepackage[T1]{fontenc}
\usepackage{ifxetex,ifluatex}
\ifnum 0\ifxetex 1\fi\ifluatex 1\fi=0 
  \usepackage[utf8]{inputenc}
\fi

\usepackage{comment}

\usepackage{graphicx}
\makeatletter
\def\maxwidth{\ifdim\Gin@nat@width>\linewidth\linewidth\else\Gin@nat@width\fi}
\def\maxheight{\ifdim\Gin@nat@height>\textheight\textheight\else\Gin@nat@height\fi}
\makeatother
\setkeys{Gin}{width=\maxwidth,height=\maxheight,keepaspectratio}

\def \citep {\cite}

\begin{document}

\excludecomment{quote}

\title{Comparison of an open-hardware electroencephalography amplifier with
medical grade device in brain-computer interface applications}


\author{
\authorname{
      Jérémy Frey
                  \unskip\sup{1, 2}
  \unskip
  }
\affiliation{\sup{1}Univ. Bordeaux, France}
\affiliation{\sup{2}Inria, France}
\email{
jeremy.frey@inria.fr}
}


\abstract{Brain-computer interfaces (BCI) are promising communication devices
between humans and machines. BCI based on non-invasive neuroimaging
techniques such as electroencephalography (EEG) have many applications,
however the dissemination of the technology is limited, in part because
of the price of the hardware. In this paper we compare side by side two
EEG amplifiers, the consumer grade OpenBCI and the medical grade g.tec
g.USBamp. For this purpose, we employed an original montage, based on
the simultaneous recording of the same set of electrodes. Two set of
recordings were performed. During the first experiment a simple adapter
with a direct connection between the amplifiers and the electrodes was
used. Then, in a second experiment, we attempted to discard any possible
interference that one amplifier could cause to the other by adding
``ideal'' diodes to the adapter. Both spectral and temporal features
were tested -- the former with a workload monitoring task, the latter
with an visual P300 speller task. Overall, the results suggest that the
OpenBCI board -- or a similar solution based on the Texas Instrument
ADS1299 chip -- could be an effective alternative to traditional EEG
devices. Even though a medical grade equipment still outperforms the
OpenBCI, the latter gives very close EEG readings, resulting in practice
in a classification accuracy that may be suitable for popularizing BCI
uses.}

\onecolumn

\maketitle

\normalsize

\vfill

\section{INTRODUCTION}\label{introduction}

\noindent Brain-computer interfaces (BCI) are communication devices
between humans and machines that rely only on brain activity (i.e.~no
muscular input) to issue commands or to monitor states
\citep{Wolpaw2002}. BCI is an emerging research area in Human-Computer
Interaction that offers new opportunities for interaction, beyond
standard input devices \citep{Tan2010}. In order to account for brain
activity, portable and non invasive neuroimaging techniques are most
commonly used, such as electroencephalography -- EEG, which measures
electrical current onto the scalp. The ``interface'' term covers many
different areas of applications, for people with or without
disabilities. However, while an increasing number of systems are being
developed, from BCI aimed at controlling a cursor \citep{Wolpaw2002} to
adaptive systems \citep{Zander2011}, more often than not the use of the
technology is limited.

The price of the hardware is one of the main reasons that prevents the
dissemination of non invasive BCI. Recently, more affordable EEG
amplifiers appeared on the market, that could solve this issue. Among
them, the OpenBCI board \footnote{\url{http://www.openbci.com/}} claims
to bring BCI to the many. Enthusiasts and laboratories have started to
use this board, but the quality of the recordings and the reliability of
the resulting systems have yet to be assessed. In this study, we compare
side by side the OpenBCI board with the g.tec g.USBamp
amplifier\footnote{\url{http://www.gtec.at/}}, a device commonly used in
BCI research. The price tag of the g.tec solution is around 20 thousands
euros, 25 times more expensive than the 800 euros of 16 channels version
of the OpenBCI board. Both OpenBCI and g.USBamp amplifiers can record up
to 16 electrodes. This number of channels is sufficient to setup various
BCI. We compared OpenBCI with g.USBamp for, on the one hand, a P300
speller application and, on the other hand, a workload monitoring
application. Doing so, we could study respectively temporal and spectral
features.

Note that here the question is not to assess which amplifier is the best
device \emph{per se}. Instead, we investigate if in a context of popular
interactions -- a narrow scope compared to the possibilities that offers
the g.USBamp -- it is conceivable for researchers from the field or
(well equipped) enthusiasts to make the leap. To which extend should we
employ devices coming from the ``DIY'' (``Do It Yourself'') community
for actual BCI applications?

To answers this question, we adopted an approach somewhat different to
what exists in the literature. Many papers deal with the comparison of
electrodes, e.g.~wet \emph{vs} dry, with or without a conductive
solution. To do so, authors try to optimize the placement of both sets
of sensors in order to get measures that originate from the same spots.
However, no matter their efforts they could not merge sensors, and even
clever montages, with electrodes of one sort positioned between
electrodes of the other sort \citep{Tautan2013}, are not ideal. It will
produce a slight offset, hence a slight inaccuracy. Another alternative
is to make separate measures by repeating the recordings with each
system \citep{Nijboer2015}, but once again the conditions could not be
exactly the same.

In the present study we do not attempt to assess the quality of
electrodes, but the behavior of amplifiers that are attached to them.
Not a whole system, \emph{only} the amplifiers. Therefore, we would not
mind using the \emph{same} electrodes during simultaneous recordings.
This setup would ensure that the signal coming in each amplifier's
inputs is exactly the same, avoiding any bias regarding the source of
the measures.

We made that possible by crafting a dedicated adapter, one that
basically splits in two the electrodes' wires. Such parallel measurement
works because the amplifiers have high impedance circuits, that is to
say that they are designed to not draw any amount of current from their
source. As such, when one amplifier is connected, the readings of the
other stay the same. Of course an infinite impedance cannot be achieved,
and no matter the precautions this setup may cause a very slight
difference compared to separate recordings. This is why in a second time
we added to our adapter a circuit that prevents any interference between
the two amplifiers, using ideal diodes to block current flows in one
direction.

Using two different BCI applications, we investigated two types of EEG
features. A task assessing workload aimed at assessing spectral
information, and an oddball task sought temporal information. For each
amplifier we measured the performance of a classifier based on those
recordings, and additionally we compared both by correlating the signals
that they recorded. No matter the financial aspects, the qualities of
the g.USBamp amplifier make it the perfect baseline to gauge new
challengers. This is also true for the electrodes developed by its
manufacturer; in this study we are using g.tec wet and active
(pre-amplified) electrodes.

\section{FIRST EXPERIMENT: DIRECT
CONNECTIONS}\label{first-experiment-direct-connections}

\subsection{Experimental setup}\label{experimental-setup}

\begin{figure}
\centering
\includegraphics[]{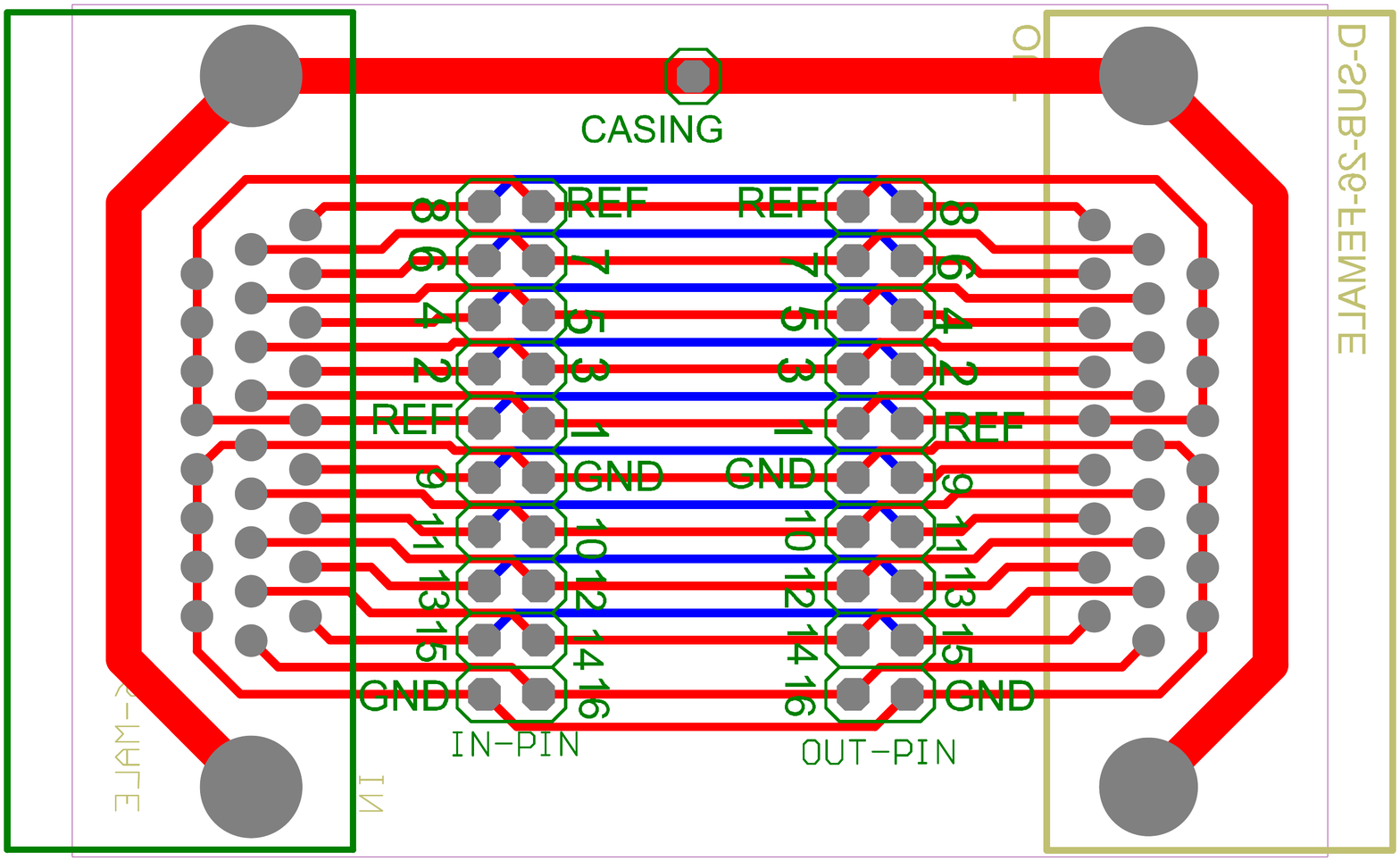}
\caption{Schematics of the direct adapter between electrodes and
amplifiers.}\label{fig:match-pcb-withoutAOP}
\end{figure}

We acquired 16 EEG channels using the active g.Ladybird electrodes from
g.tec. In this system, the electrodes are attached to a box that powers
their electrical components and retrieves the signal; the g.GAMMAbox.
After studying the wiring of the g.GAMMAbox, we designed a printed
circuit board (PCB) to connect both amplifiers. Our adapter plugs on one
end to the D-sub 26 connector of the g.GAMMAbox. Thanks to a pinout
composed of 2.54mm connectors that gave access to all the channels (16
EEG + reference + ground), we attached the OpenBCI board to the adapter
-- ground set to ``bias'' pin. On the other end of the adapter there was
a D-sub 26 female connector, onto which we could plug the g.USBamp
amplifier as if it were the regular end of the g.GAMMAbox. The
schematics of the 2 layers PCB is presented in
Figure~\ref{fig:match-pcb-withoutAOP}.

The EEG channels were positioned according to the 10-20 system at AFz,
Fz, FCz, C3, C1, Cz, C2, C4, CPz, P3, Pz, P4, POz, O1, Oz and O2 --
ground at FPz, reference on the left earlobe. Since the measures between
both amplifiers were identical, only one recording session occurred,
with one participant -- there were no factors to counterbalance with
repeated measures. The signals of both amplifier were acquired using
OpenViBE 1.0.1\footnote{\url{http://openvibe.inria.fr/}}; at 512Hz
sampling rate for the g.USBamp and 125Hz sampling rate for the OpenBCI
board.

The spectral features were investigated with a workload monitoring
application, i.e.~a BCI that is able to discriminate between several
levels of mental effort. The system was trained using the N-back task, a
well-known task to induce workload by playing on memory load and time
pressure (see, e.g. \citep{Muhl2014}). The protocol we used was similar
to \citep{Muhl2014}, there were 360 trials presented during 6 blocks of
alternate difficulty levels (0-back \emph{vs} 2-back conditions). The
recording session lasted approximately 12 minutes.

The temporal features were investigated using an oddball task directly
implemented within OpenViBE with a visual P300 speller. P300 spellers
are well-established BCI applications during which letters that randomly
flash on the screen can be used to spell words with the sole brain
activity. Indeed, when the letter that the user wants to spell flashes,
a particular event-related potential (ERP) arises within the EEG, which
possess a positive ``peak'' around t=300 ms after the stimulus onset.
This is commonly referred to as the ``oddball paradigm'' since the
occurrence of rare stimuli is used to elicit brain responses. During the
recordings a matrix of 6 by 6 letters and digits was displayed in full
screen on a 24-inch display. Only the calibration session occurred,
during which one need to focus one's attention on a predefined sequence
of letters. 32 letters composing a pangram were mentally ``spelled''
this way. The sentence was, without spaces, ``pack my box with five
dozen liquor jugs''. Letters were flashing for 0.2s. There were 24
flashes per letter (12 times the row, 12 times the column), hence due to
the matrix disposition there were in total 4608 trials, among which 768
were targets -- ``odd'' trials, i.e.~the letters of the target sentence
were flashing. The recording session lasted approximately 30 minutes.

The acquisition of both amplifiers' signals and the P300 application
occurred within the same OpenViBE scenario (script). The recordings of
each amplifier were synchronized with the appropriate events and
exported in separate GDF files for later analyses. There was also only
one scenario involved in the synchronization of all signals and events
in the case of the N-back task; stimulation from the python script
supporting this latter task were retrieved using the LSL protocol, a
network protocol dedicated to physiological recordings which ensures
accurate timings\footnote{\url{https://github.com/sccn/labstreaminglayer}}.

\subsection{Signal processing}\label{signal-processing}

Two kinds of analyses were performed, using standard BCI signal
processing pipelines. One aimed at assessing if and how the amplifiers
differ in practice, when used for classification. The second then looked
at the correlation between the acquired signals.

\subsubsection{Classification}\label{classification}

The signal processing of the data acquired during the N-back task is
analogous to \citep{Muhl2014}, i.e.~2s time windows, 5 frequency bands
-- delta (1-3 Hz), theta (4-6 Hz), alpha (7-13 Hz), beta (14-25 Hz) and
gamma (26-40 Hz) -- and spatial filters. We used common spatial patterns
spatial filters to reduce the 16 channels to 6 ``virtual'' channels more
discriminant between the workload conditions -- see \citep{Ramoser2000}.
Additionally, we also tested a 3 frequency bands version of our
pipeline, that consider only the lower frequencies, less prone to
muscular artifacts -- delta, theta and alpha.

Concerning the oddball task, we band-passed the signal between 0.5Hz and
40Hz, downsampled it by a factor 32 using the ``decimate'' Matlab
function -- by a factor 8 for OpenBCI because of the reduced sampling
rate --, and applied regularized Eigen Fisher spatial filters -- a
spatial filter specifically designed for ERPs classification
\citep{Hoffmann2006} -- to reduce channels' dimension from 16 to 5. We
used 1s time windows after stimuli onsets -- letters' flashes -- to
epoch (``slice'') our signal. However, in order to prevent data to
overlap between consecutive stimuli due to the rapid pace of the
flashes, after a first pass of epoching we discarded overlapping time
windows from further analyses. This ensured that no part of the signal
could be seen twice by the classifier between the training phase and the
testing phase and bias the accuracy. The procedure was automatic, the
first non-overlapping epoch in order of appearance being kept. As a
result, in the end we obtained 48 target trials and 240 distractor
trials for classification, identical between the g.USBamp and the
OpenBCI recordings.

Both for the workload and the P300 speller tasks, we used shrinkage LDA
(linear discriminant analysis) for classification \citep{Ledoit2004}. To
assess the classifiers' performance on the calibration data, we used
4-fold cross-validation. I.e. we split the collected data into 4 parts
of equal size, used 3 parts to calibrate the classifiers and tested the
resulting classifiers on the unseen data from the remaining part. This
process occurred 3 more times so that in the end each part was used once
as test data. Finally, we averaged the obtained classification
accuracies. The accuracy was measured using the area under the
receiver-operating characteristic curve (AUROCC). The AUROCC is a metric
that is robust against unbalanced classes, as it is the case with
oddball tasks. A score of ``1'' means a perfect classification, a score
of ``0.5'' is chance. In order to make statistical comparisons between
both amplifiers for each type of features that we studied, we ran 10
times the analyses -- the trials were selected randomly for
cross-validation.

\subsubsection{Correlations}\label{correlations}

We compared, on the one hand, the frequency spectra associated to the
different workload conditions and, on the other hand, the time course of
the ERP that were caused by the flashing target letters. To do so, we
used Pearson correlations, on par with the literature for similar
analyses -- e.g. \citep{Zander2011b}. In order to ensure a 1-to-1
correspondence between our sets of data, the recordings from the
g.USBamp were downsampled to 125Hz -- same sampling rate as for the
OpenBCI -- using the ``resample'' function from Matlab R2014a signal
processing toolbox.

Concerning the workload task, we first aggregated the 2s time-windows
corresponding to each condition (0-back and 2-back). Then we used the
``spectopo'' function of the EEGLAB toolbox (version 13.4.4b) to compute
the grand average power spectral between 1Hz and 40Hz, for each channel.
The output of the function was then passed on to R (version 3.0.2) to
compute correlations through the ``rcorr'' function from the ``Hmisc''
package.

For the oddball task, we first band-passed the signals between 1Hz and
8Hz -- the approximate frequency band used for classification. Then we
extracted time epochs starting 0.5s prior to the flashing of the target
letters and ending 1s after stimuli onset. Contrary to what occurred for
classification, we did not prune overlapping epochs in the oddball task
when we compute the averaged ERP -- there was no bias that could have
been induced here. Finally, we averaged the ERP per channel before
exporting the time points to the R environment.

\subsection{Results}\label{results}

\begin{figure*}
\centering
\includegraphics[]{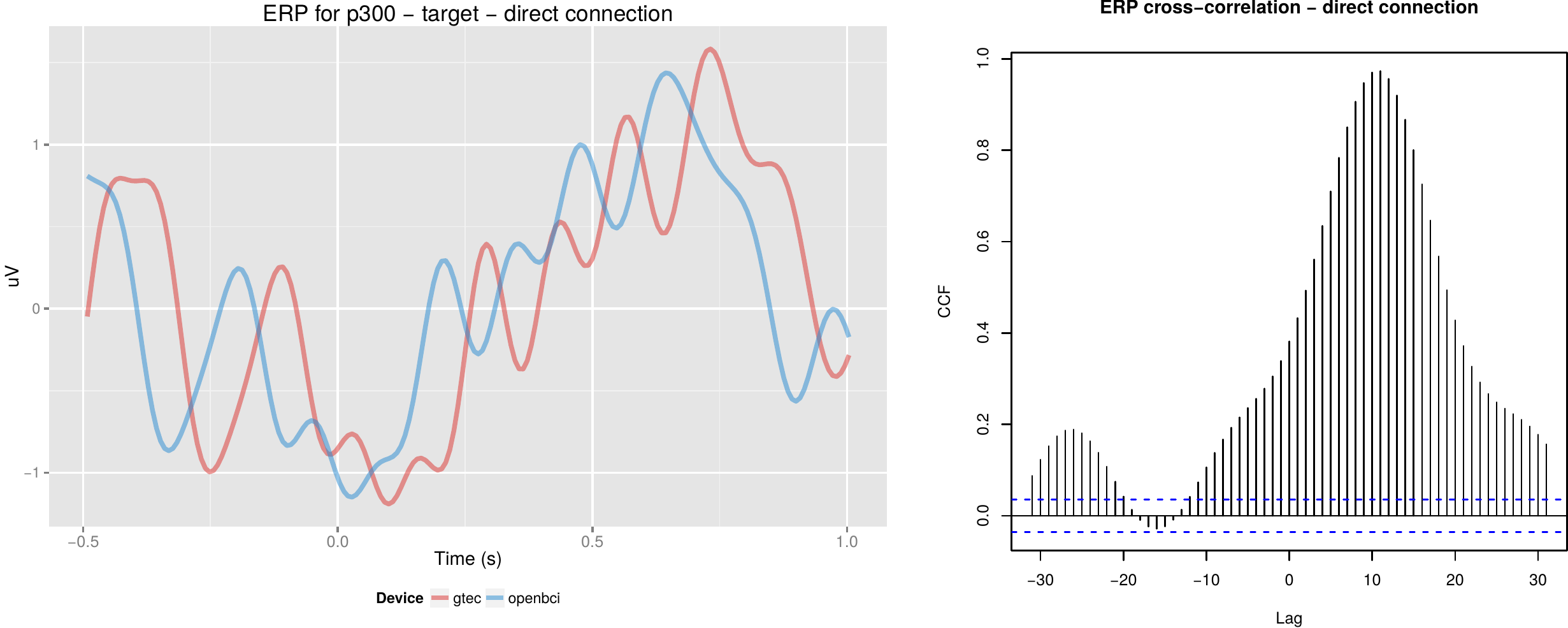}
\caption{\emph{Left}: Averaged ERP across channels of the target trials
during the oddball task, before time shift correction. \emph{Right}:
Cross-correlation between the amplifiers. The computed lag of 11 data
points corresponds to 88ms. (Direct
connection.)}\label{fig:match-direct-ccf}
\end{figure*}

\begin{figure*}
\centering
\subfloat[\label{fig:match-direct-p300-all}]{\includegraphics[width=0.500\hsize]{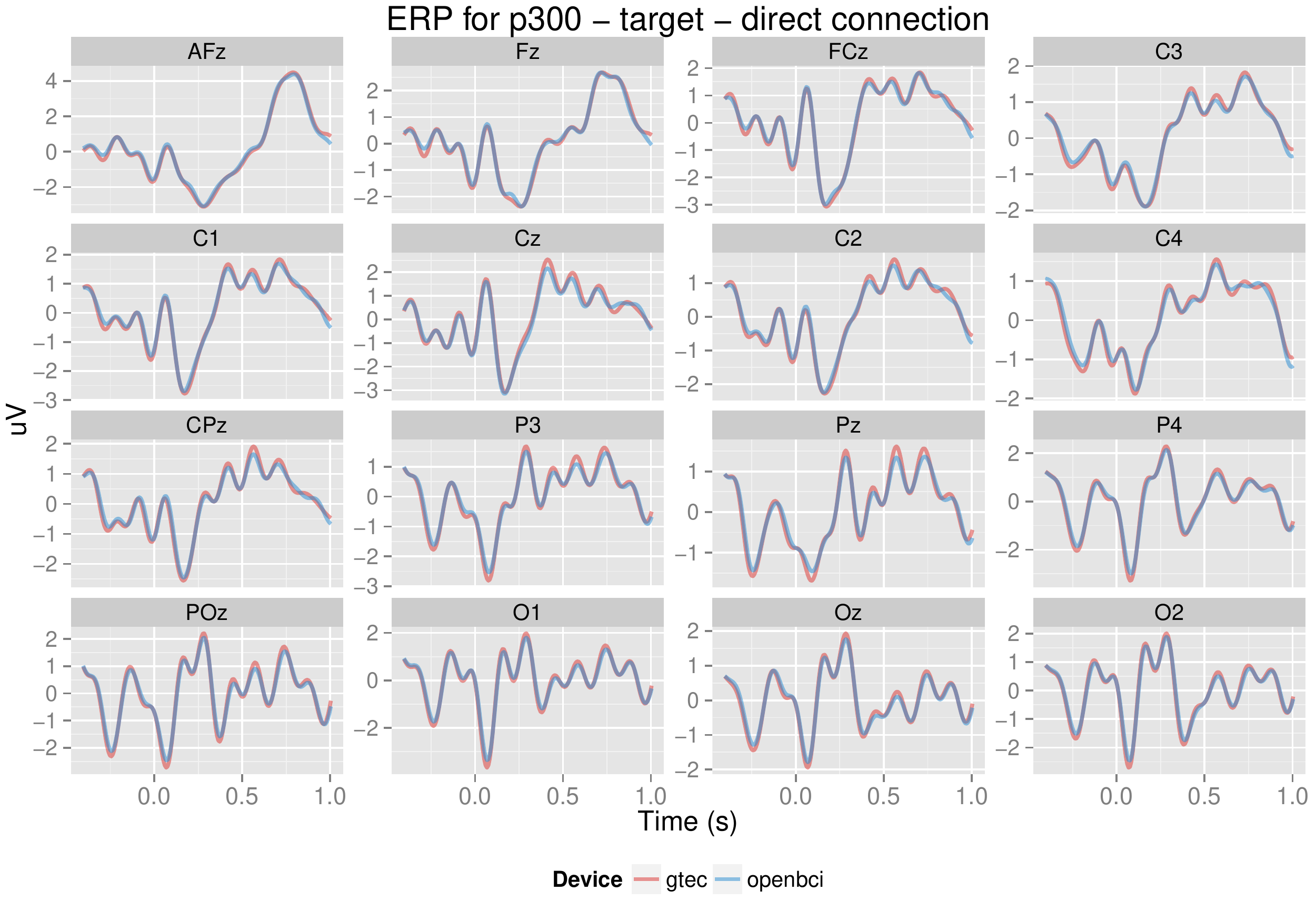}}
\\
\subfloat[\label{fig:match-direct-0back-all}]{\includegraphics[width=0.500\hsize]{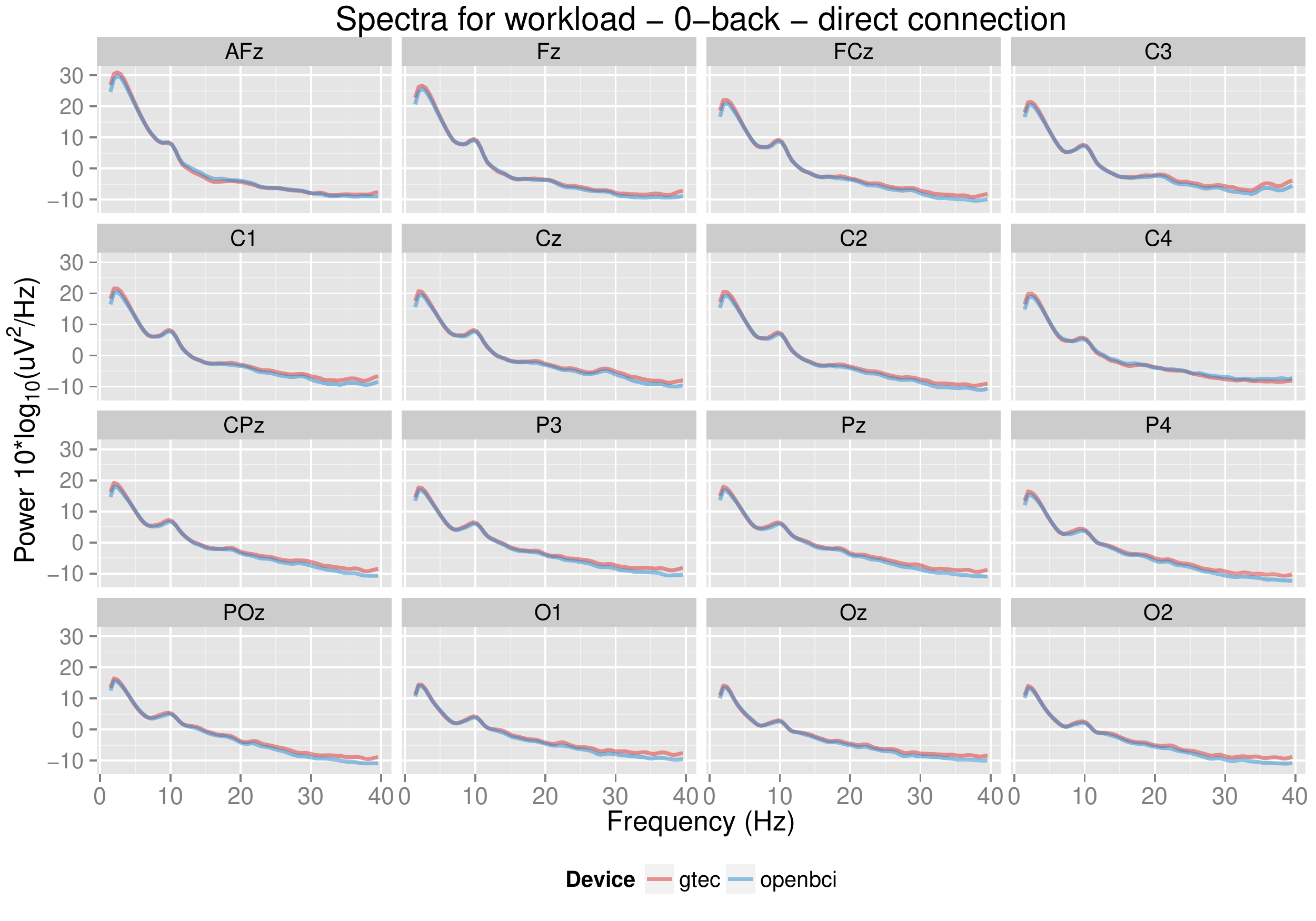}}
\subfloat[\label{fig:match-direct-2back-all}]{\includegraphics[width=0.500\hsize]{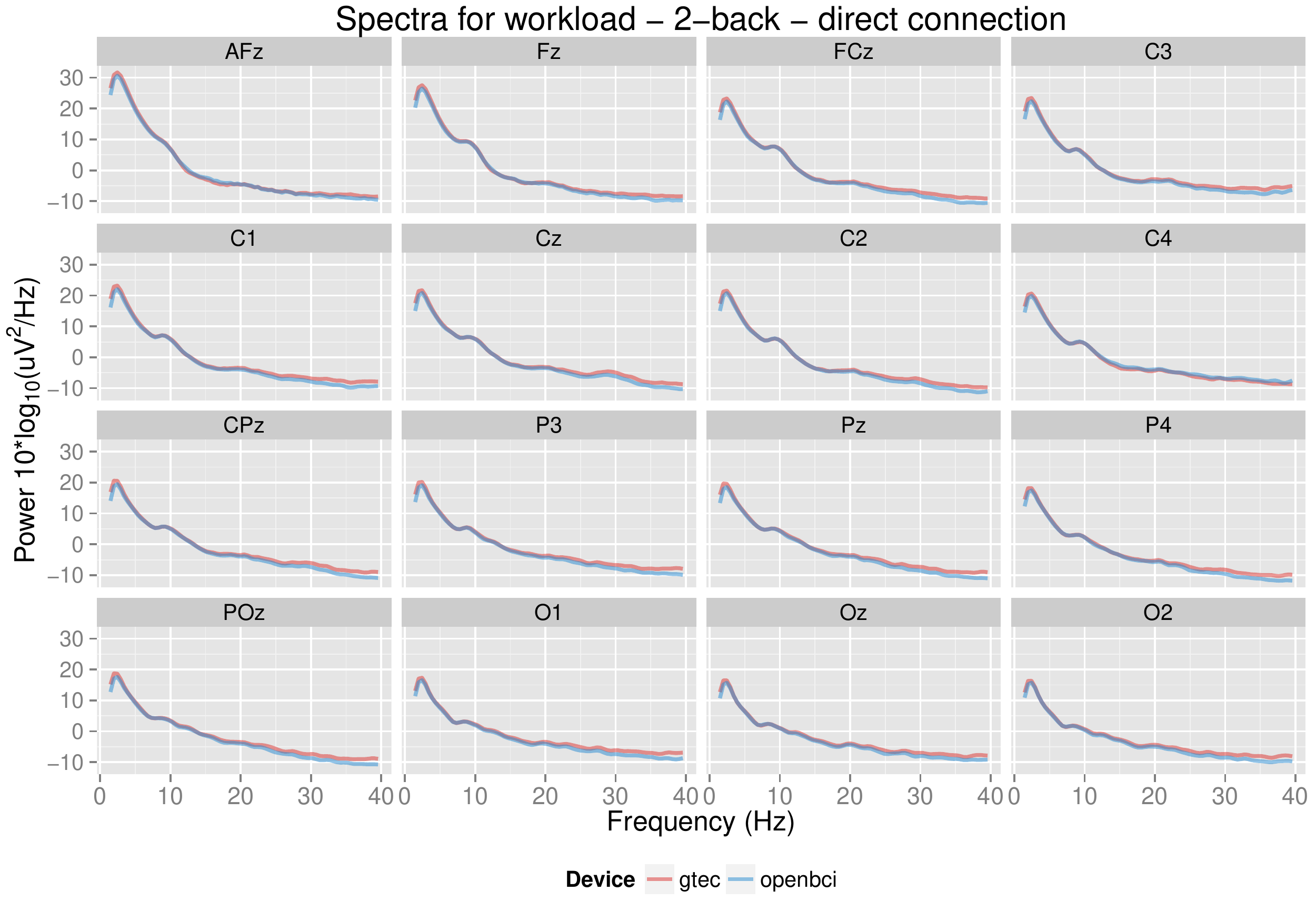}}
\caption{Direct connection: averaged ERP for the target trials of the
oddball task (\emph{a}), averaged spectra for the 0-back (\emph{b}) and
2-back (\emph{c}) trials of the N-bak task.}\label{fig:match-direct-all}
\end{figure*}

\subsubsection{Classification}\label{classification-1}

The results regarding classification accuracy are presented in
Table~\ref{tab:match-direct-class}, with the AUROCC scores for each one
of the 10 repetitions, for both amplifiers and both tasks -- including
the 3 and 5 frequency bands pipeline for workload.

We tested for significance using Wilcoxon signed-rank tests. There was a
significant difference between amplifiers for the P300 tasks
(p~\textless{}~0.01). The AUROCC mean score for the g.USBamp was 0.961
\emph{vs} 0.918 for the OpenBCI. There were however no significance but
tendencies concerning the workload task, with mean AUROCC scores between
0.85 and 0.86 for the 3 bands pipeline (p = 0.095) and between 0.89 and
0.90 for the 5 bands task (p = 0.079) -- see
Table~\ref{tab:match-direct-class} for details.

\begin{table*}
\centering
\begin{tabular}{llllllllllllll}
\toprule\addlinespace
\textbf{Condition} & \textbf{Amplifier} & \textbf{1} & \textbf{2} &
\textbf{3} & \textbf{4} & \textbf{5} & \textbf{6} & \textbf{7} &
\textbf{8} & \textbf{9} & \textbf{10} & \textbf{\emph{Mean}} &
\textbf{\emph{SD}}\tabularnewline
\midrule
P300 & g.USBamp & 0.96 & 0.96 & 0.96 & 0.96 & 0.96 & 0.96 & 0.96 & 0.96
& 0.96 & 0.97 & \emph{0.961} & \emph{0.003}\tabularnewline
& OpenBCI & 0.92 & 0.92 & 0.91 & 0.91 & 0.92 & 0.92 & 0.93 & 0.92 & 0.92
& 0.91 & \emph{0.918} & \emph{0.006}\tabularnewline
WL 3b & g.USBamp & 0.85 & 0.85 & 0.85 & 0.86 & 0.85 & 0.87 & 0.87 & 0.87
& 0.87 & 0.86 & \emph{0.860} & \emph{0.009}\tabularnewline
& OpenBCI & 0.86 & 0.86 & 0.85 & 0.86 & 0.84 & 0.86 & 0.85 & 0.85 & 0.85
& 0.85 & \emph{0.853} & \emph{0.007}\tabularnewline
WL 5b & g.USBamp & 0.90 & 0.89 & 0.90 & 0.89 & 0.91 & 0.90 & 0.89 & 0.90
& 0.89 & 0.90 & \emph{0.897} & \emph{0.007}\tabularnewline
& OpenBCI & 0.91 & 0.89 & 0.87 & 0.88 & 0.89 & 0.88 & 0.89 & 0.89 & 0.89
& 0.90 & \emph{0.889} & \emph{0.011}\tabularnewline
\bottomrule
\end{tabular}
\caption{Classification accuracy (AUROCC scores) for the P300 and
workload tasks studied during the first experiment -- direct connection
between the electrodes and the amplifiers. The 4-fold cross validations
were repeated 10 times. Two pipelines are presented for the workload: 3
frequency bands (``WL 3b'', $\delta + \theta + \alpha$) as well as 5
frequency bands pipeline (``WL 5b'',
$\delta + \theta + \alpha + \beta + \gamma$). Significance was tested
using Wilcoxon signed-rank tests.}\label{tab:match-direct-class}
\end{table*}

\begin{table*}
\centering
\begin{tabular}{llllllllll}
\toprule\addlinespace
& \textbf{AFz} & \textbf{Fz} & \textbf{FCz} & \textbf{C3} & \textbf{C1}
& \textbf{Cz} & \textbf{C2} & \textbf{C4} & \textbf{CPz}\tabularnewline
\midrule
P300 target & 0.998 & 0.997 & 0.997 & 0.997 & 0.997 & 0.994 & 0.996 &
0.992 & 0.994\tabularnewline
Workload 0-back & 0.999 & 0.999 & 0.998 & 0.998 & 0.998 & 0.998 & 0.998
& 0.999 & 0.998\tabularnewline
Workload 2-back & 0.999 & 0.999 & 0.998 & 0.998 & 0.998 & 0.998 & 0.998
& 0.999 & 0.997\tabularnewline
& \textbf{P3} & \textbf{Pz} & \textbf{P4} & \textbf{POz} & \textbf{O1} &
\textbf{Oz} & \textbf{O2} & \textbf{\emph{Mean}} &
\textbf{\emph{SD}}\tabularnewline
P300 target & 0.995 & 0.994 & 0.996 & 0.996 & 0.996 & 0.994 & 0.995 &
\emph{0.9965} & \emph{0.0015}\tabularnewline
Workload 0-back & 0.998 & 0.998 & 0.998 & 0.999 & 0.998 & 0.998 & 0.998
& \emph{0.9983} & \emph{0.0003}\tabularnewline
Workload 2-back & 0.998 & 0.998 & 0.998 & 0.998 & 0.997 & 0.998 & 0.998
& \emph{0.9979} & \emph{0.0005}\tabularnewline
\bottomrule
\end{tabular}
\caption{Pearson correlation R scores between g.USBamp and OpenBCI
recordings at the 16 different electrode locations with a direct
connection. The ``P300 target'' condition corresponds to temporal
features (ERP averaged across trials) and the workload conditions to
spectral features.}\label{tab:match-direct-correlation}
\end{table*}

\subsubsection{Correlations}\label{correlations-1}

When we first analyzed our data to seek correlations regarding the
oddball tasks, we realized that a shift occurred during the recordings,
as denoted in Figure~\ref{fig:match-direct-ccf} by the grand average of
the ERP for target trials across channels. This may have been caused by
a software issue (see Discussion). In order to correct the shift and
conduct proper comparisons between both amplifiers' measures, we used a
cross-correlation to estimate the time shift, using the ``ccf'' function
from the R ``stats'' package. We found a delay of 88ms between the two
signals -- 11 data points at 125Hz, see
Figure~\ref{fig:match-direct-ccf}.

In Figure~\ref{fig:match-direct-p300-all}, the averaged ERP were shifted
by as much for each channel. Corresponding Pearson correlation R scores,
that were computed using the ``rcorr'' function, are presented in
Table~\ref{tab:match-direct-correlation}. The mean R score is 0.9965 and
is statistically significant (p~\textless{}~0.001).

There was also a significant correlation (p~\textless{}~0.001) for the
spectral features, with a mean R score of 0.9983 for the 0-back
condition and 0.9979 the 2-back condition (see
Table~\ref{tab:match-direct-correlation} for details). Among the brain
signals patterns that could be expected during the completion of a
difficult task, the decrease nearby the alpha frequency band during the
2-task condition can be observed within per-channel spectra presented in
Figures~\ref{fig:match-direct-0back-all}
and~\ref{fig:match-direct-2back-all}. Note that we did not correct time
shifts prior to workload analyses due to the nature of the features --
i.e.~spectral and not temporal.

\subsection{Discussion}\label{discussion}

The correlation between both temporal and spectral features tends so
show that the signals acquired by the g.USBamp and the OpenBCI are, if
not identical, very closely related. For every condition and channel
tested, the Pearson R score was greater than 0.99.

There were however more dissimilarities in the classification accuracy
obtained during the corresponding tasks. While there were hardly a
difference between the AUROCC scores computed from both amplifiers with
the N-back tasks, the g.USBamp performed significantly better than the
OpenBCI during the P300 speller task. The time shift observed afterwards
between the two amplifiers may partially explain this difference.
Indeed, the detection of ERP is particularly sensitive to signals'
latency, and a shift between events' timestamp and signal's acquisition
could result in such degradation of performance when temporal features
are involved.

The radio transmission between the wireless OpenBCI board and the dongle
plugged to the computer may be one of the cause of the situation. The
problem could also originate from the software. As a matter of fact, the
OpenViBE acquisition driver of the OpenBCI board was released no so long
before our experiment, and was still labelled as ``unstable'' as for
version 1.0.1 of the software. One ``oddity'' that may further highlight
the youth of OpenBCI software integration: we realized during our
analysis that the recorded signals were completely inverted on the Y
axis. The voltage reported by the board were the opposite of what
g.USBamp was claiming. Since on numerous occasions we acknowledged the
accuracy of g.tec devices readings, it is the OpenBCI's signals that we
inverted back to ``normal'' prior to correlation analyses.

Beside time shifts issues, as mentioned during the introduction we
needed to strengthen those first insights by discarding the eventuality
that both EEG signals may have influence each other due to the direct
wiring with the electrodes.

\section{SECOND EXPERIMENT: ISOLATED
CONNECTIONS}\label{second-experiment-isolated-connections}

\noindent The second set of recordings is very similar to the what was
described during the first study. The second experiment only differs by
the nature of the adapter that was employed. As such we will only
discuss the changes that were made to the hardware and quickly dive into
the results.

\subsection{Ideal adapter}\label{ideal-adapter}

We modified the adapter that connects the amplifiers to the g.GAMMAbox
-- and by extent to the EEG electrodes. Instead of a direct connection
between each amplifier's inputs and the EEG channels, we interposed
``ideal'' (or ``super'') diodes on the branches of the ``Y'' wiring.

Diodes are electrical components that let the current flow in only one
direction, the ``forward'' direction. Hence, this type of montage
ensures that no current could travel directly from one amplifier to the
other, contaminating the recordings. However, regular diodes cause a
voltage drop. The voltage drop varies depending on the materials used
for their construction, but it is at least 0.3V. Meaning that if the
current coming in the forward direction is lesser than 0.3V, no signal
will pass through. 0.3V is an order of magnitude superior to the range
of EEG signal -- approximately a thousand time, therefore regular diode
could not be used.

\begin{figure}
\centering
\includegraphics[]{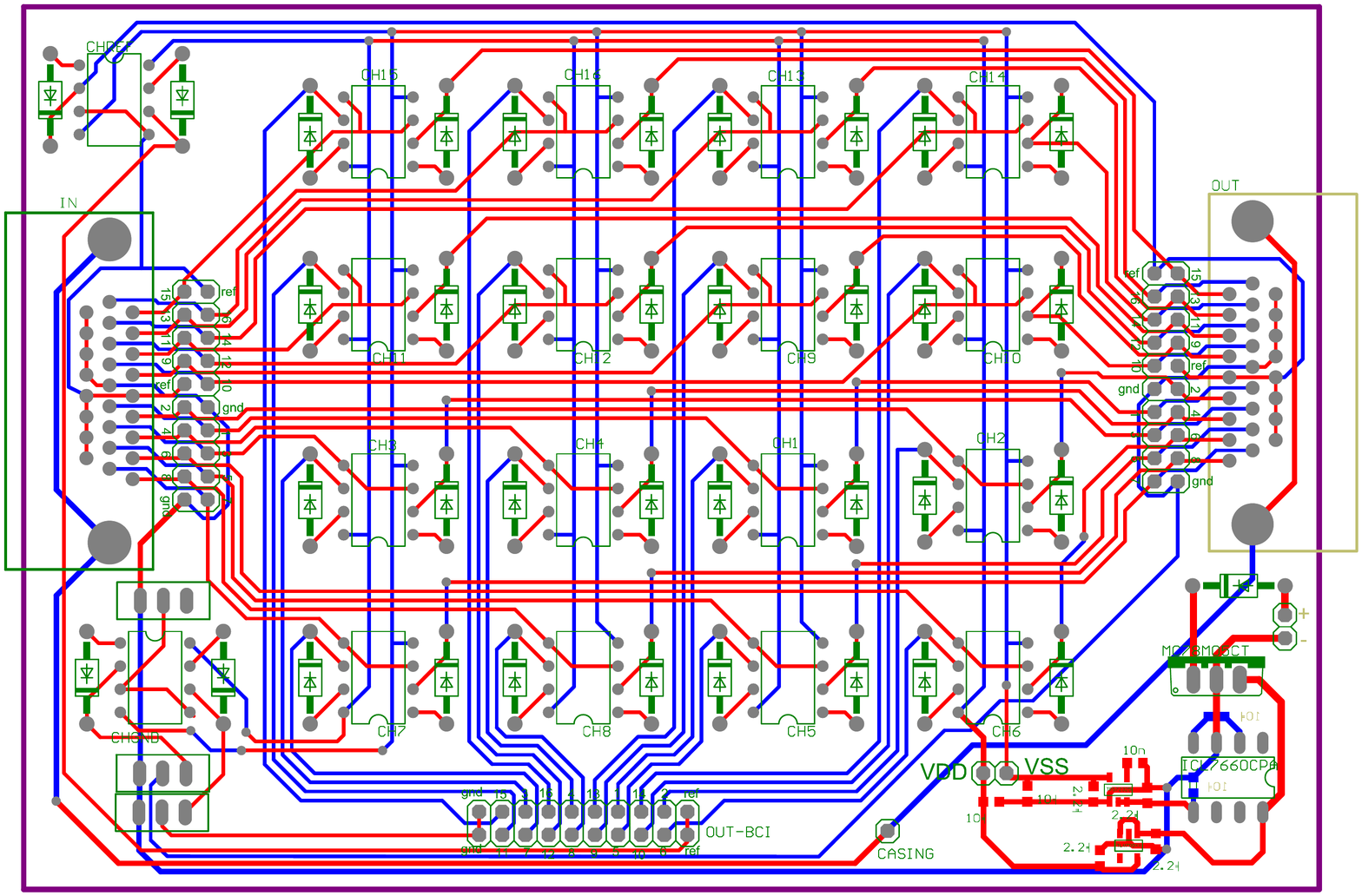}
\caption{Schematics of the adapter with the ideal diodes montage. Note
that there is a set of ideal diodes on the ground channel, but they were
shorted with a jumper during our
experiment.}\label{fig:match-pcb-withAOP}
\end{figure}

To circumvent this problem, we utilized a particular montage that
involved operational amplifiers (op-amp). Op-amps are components widely
used in electrical circuits, acting as sorts of ``building blocks''.
Notably, in combination with a regular diode, one could use a precision
rectifier configuration to obtain an ``ideal'' diode. This particular
montage is also known as a ``super'' diode, since there will always be a
slight voltage drop, but in this case, thanks to the gain of the op-amp,
it becomes negligible.

We mounted 36 of such ideal diodes on the adapter. One on each end of
the ``Y'' section associated to the 16 EEG channels, plus 2 for the
reference. Due to the nature of the electrical recordings, only the
ground was left without such circuit. We utilized Texas Instrument
op-amps, model TLC2272ACPE4. The TLC227xA series are more indicated for
precision application, and with 2 op-amps per chip we could limit the
overall size of the adapter. The operational amplifiers were powered by
an external circuit with regulated -2.5 / +2.5 voltage. The schematics
of the adapter -- also a 2 layers 2 layers PCB -- is presented in
Figure~\ref{fig:match-pcb-withAOP}. The ideal diodes montage, placed
before amplifiers' inputs, prevented any current to flow in reverse
direction from either amplifier to the adapter; it ensured that one set
of recordings would not bias the other. One recording session occurred
for each application and each condition.

\subsection{Results}\label{results-1}

The signal processing and the analyses were strictly identical to the
first experiment detailed above, refer to the previous section for
related information.

\subsubsection{Classification}\label{classification-2}

As with the first study, the results regarding classification accuracy
are presented in Table~\ref{tab:match-split-class}, with the AUROCC
scores for each one of the 10 repetitions, for both amplifiers and both
tasks -- including the 3 and 5 frequency bands pipeline for workload. We
tested for significance using Wilcoxon signed-rank tests. No matter the
task there was no significant difference, although the 5\% threshold was
nearly reached for spectral features -- p-value was 0.157 for the P300
task, \emph{0.051} for the 3 bands version of the workload pipeline and
0.286 for the 5 bands version.

\subsubsection{Correlations}\label{correlations-2}

\begin{figure*}
\centering
\includegraphics[]{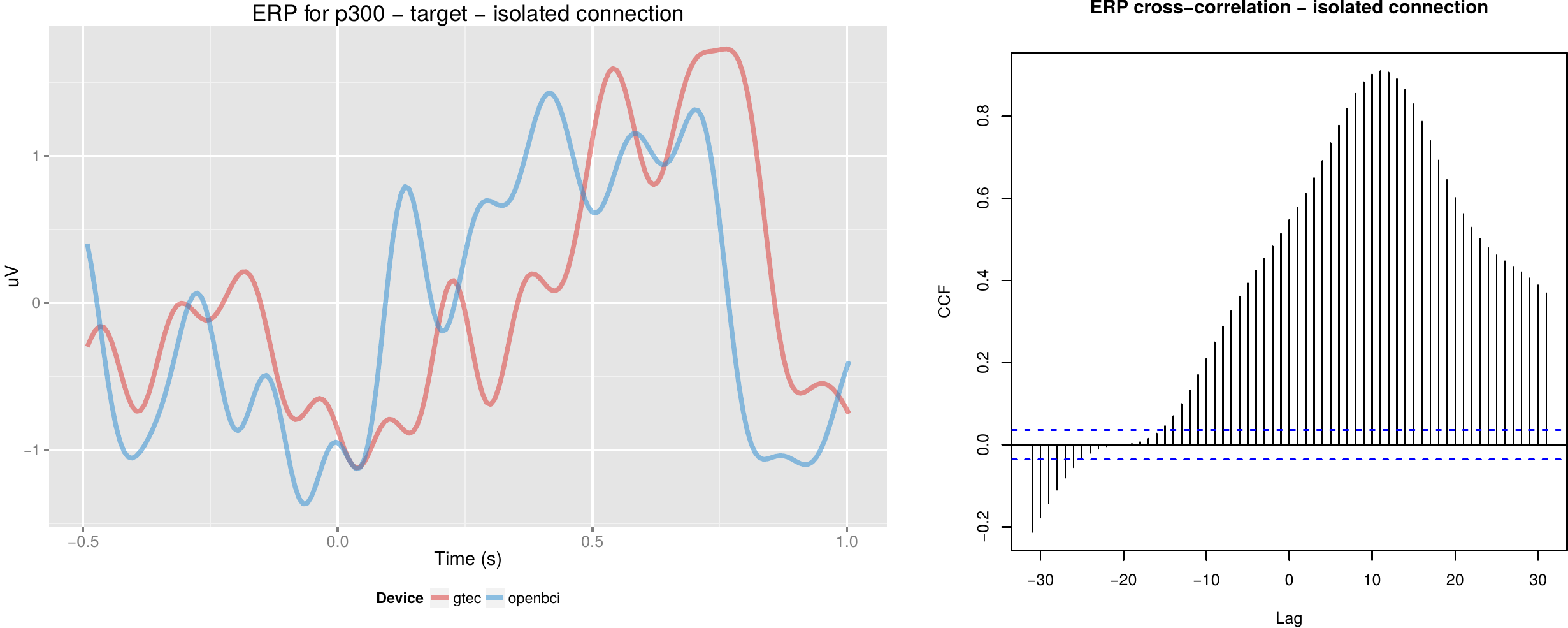}
\caption{\emph{Left}: Averaged ERP across channels of the target trials
during the oddball task, before time shift correction. \emph{Right}:
Cross-correlation between the amplifiers. The computed lag of 11 data
points corresponds to 88ms. (Isolated
connection.)}\label{fig:match-split-ccf}
\end{figure*}

\begin{figure*}
\centering
\subfloat[\label{fig:match-split-p300-all}]{\includegraphics[width=0.500\hsize]{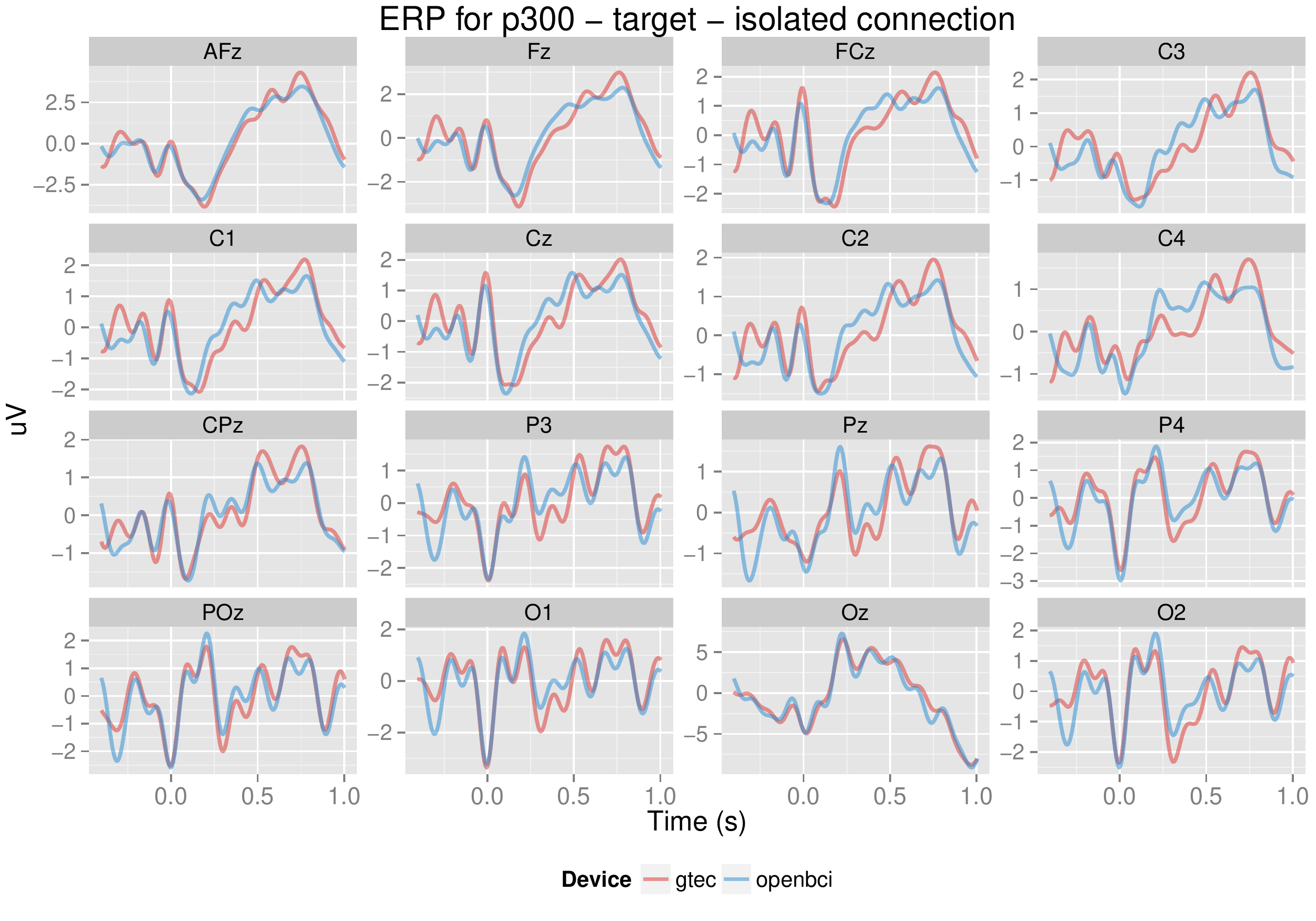}}
\\
\subfloat[\label{fig:match-split-0back-all}]{\includegraphics[width=0.500\hsize]{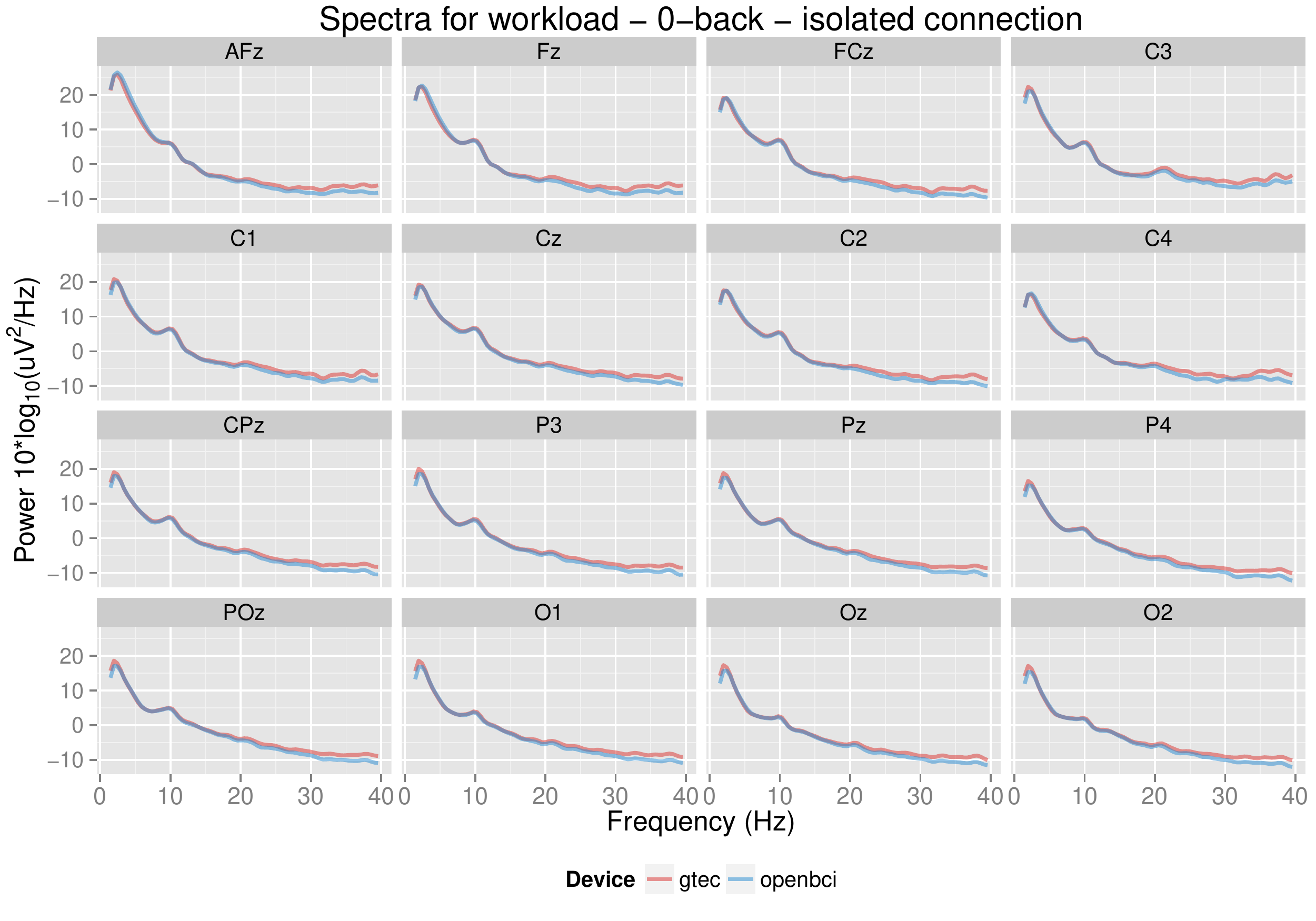}}
\subfloat[\label{fig:match-split-2back-all}]{\includegraphics[width=0.500\hsize]{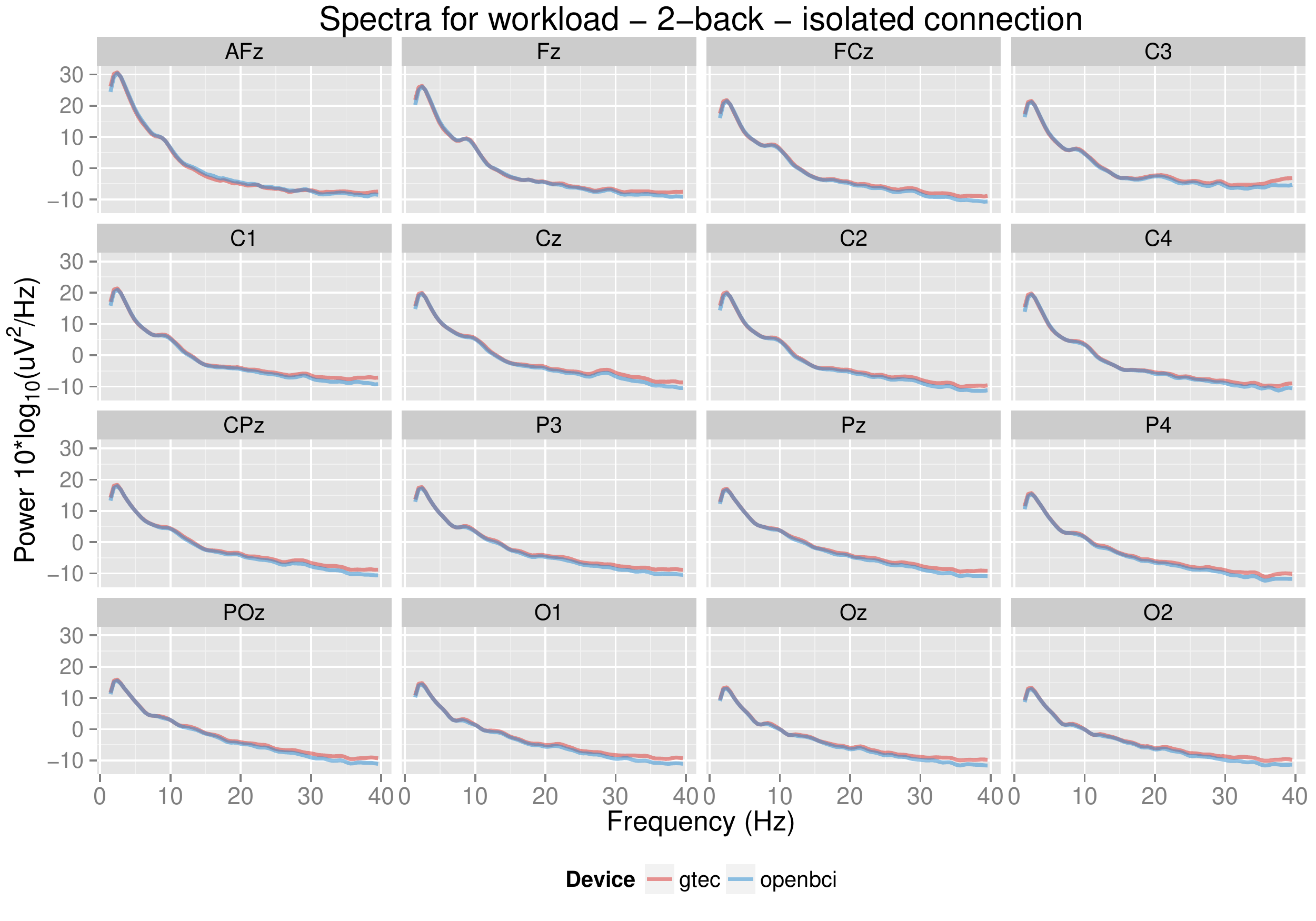}}
\caption{Isolated connection: averaged ERP for the target trials of the
oddball task (\emph{a}), averaged spectra for the 0-back (\emph{b}) and
2-back (\emph{c}) trials of the N-bak task.}\label{fig:match-split-all}
\end{figure*}

Concerning the P300 oddball task, there was a offset of 88ms as well
between the recordings of both amplifier with the isolated connection --
see Figure~\ref{fig:match-split-ccf} for the grand ERP average and the
cross correlation. The per-channel averaged ERP are plotted in
figure~\ref{fig:match-split-p300-all}. Corresponding Pearson correlation
R scores are presented in Table~\ref{tab:match-split-correlation}. The
mean R score is 0.8847 and is statistically significant
(p~\textless{}~0.001).

There was also a significant correlation (p~\textless{}~0.001) for the
spectral features, with a mean R score of 0.9976 for the 0-back
condition and 0.9987 the 2-back condition (see
Table~\ref{tab:match-split-correlation} for details). The per-channel
spectra are presented in Figures~\ref{fig:match-split-0back-all}
and~\ref{fig:match-split-2back-all}. As with the direct connection, the
band frequency changes between the 0-back and the 2-back conditions can
be observed in the spectra.

\begin{table*}
\centering
\begin{tabular}{llllllllllllll}
\toprule\addlinespace
\textbf{Condition} & \textbf{Amplifier} & \textbf{1} & \textbf{2} &
\textbf{3} & \textbf{4} & \textbf{5} & \textbf{6} & \textbf{7} &
\textbf{8} & \textbf{9} & \textbf{10} & \textbf{\emph{Mean}} &
\textbf{\emph{SD}}\tabularnewline
\midrule
P300 & g.USBamp & 0.84 & 0.82 & 0.83 & 0.83 & 0.81 & 0.85 & 0.83 & 0.84
& 0.83 & 0.84 & \emph{0.832} & \emph{0.011}\tabularnewline
& OpenBCI & 0.82 & 0.83 & 0.83 & 0.82 & 0.81 & 0.82 & 0.84 & 0.82 & 0.83
& 0.83 & \emph{0.825} & \emph{0.008}\tabularnewline
WL 3b & g.USBamp & 0.88 & 0.88 & 0.88 & 0.91 & 0.89 & 0.90 & 0.89 & 0.88
& 0.89 & 0.88 & \emph{0.888} & \emph{0.100}\tabularnewline
& OpenBCI & 0.88 & 0.88 & 0.89 & 0.89 & 0.88 & 0.89 & 0.87 & 0.88 & 0.88
& 0.87 & \emph{0.881} & \emph{0.007}\tabularnewline
WL 5b & g.USBamp & 0.92 & 0.91 & 0.92 & 0.90 & 0.91 & 0.89 & 0.90 & 0.91
& 0.91 & 0.90 & \emph{0.907} & \emph{0.009}\tabularnewline
& OpenBCI & 0.90 & 0.92 & 0.92 & 0.91 & 0.92 & 0.90 & 0.91 & 0.91 & 0.90
& 0.91 & \emph{0.910} & \emph{0.008}\tabularnewline
\bottomrule
\end{tabular}
\caption{Classification accuracy (AUROCC scores) for the P300 and
workload tasks studied during the second experiment -- isolated
connection between the electrodes and the amplifiers. The 4-fold cross
validations were repeated 10 times. Two pipelines are presented for the
workload: 3 frequency bands (``WL 3b'', $\delta + \theta + \alpha$) as
well as 5 frequency bands pipeline (``WL 5b'',
$\delta + \theta + \alpha + \beta + \gamma$). Significance was tested
using Wilcoxon signed-rank tests.}\label{tab:match-split-class}
\end{table*}

\begin{table*}
\centering
\begin{tabular}{llllllllll}
\toprule\addlinespace
& \textbf{AFz} & \textbf{Fz} & \textbf{FCz} & \textbf{C3} & \textbf{C1}
& \textbf{Cz} & \textbf{C2} & \textbf{C4} & \textbf{CPz}\tabularnewline
\midrule
P300 target & 0.976 & 0.934 & 0.892 & 0.846 & 0.872 & 0.881 & 0.838 &
0.811 & 0.912\tabularnewline
Workload 0-back & 0.999 & 0.998 & 0.998 & 0.997 & 0.998 & 0.998 & 0.998
& 0.996 & 0.998\tabularnewline
Workload 2-back & 0.999 & 0.999 & 0.999 & 0.998 & 0.998 & 0.999 & 0.999
& 0.999 & 0.999\tabularnewline
& \textbf{P3} & \textbf{Pz} & \textbf{P4} & \textbf{POz} & \textbf{O1} &
\textbf{Oz} & \textbf{O2} & \textbf{\emph{Mean}} &
\textbf{\emph{SD}}\tabularnewline
P300 target & 0.849 & 0.823 & 0.896 & 0.910 & 0.878 & 0.982 & 0.853 &
\emph{0.8847} & \emph{0.0483}\tabularnewline
Workload 0-back & 0.998 & 0.998 & 0.998 & 0.997 & 0.997 & 0.997 & 0.997
& \emph{0.9976} & \emph{0.0007}\tabularnewline
Workload 2-back & 0.999 & 0.999 & 0.999 & 0.999 & 0.999 & 0.998 & 0.998
& \emph{0.9987} & \emph{0.0004}\tabularnewline
\bottomrule
\end{tabular}
\caption{Pearson correlation R scores between g.USBamp and OpenBCI
recordings at the 16 different electrode locations with an isolated
connection. The ``P300 target'' condition corresponds to temporal
features (ERP averaged across trials) and the workload conditions to
spectral features.}\label{tab:match-split-correlation}
\end{table*}

\subsection{Discussion}\label{discussion-1}

The results with the isolated connections are not that different from
what was obtained during the first experiment. This would suggest that
directly connecting two high impedance amplifiers to the same EEG
electrodes could be a viable montage for a side-by-side comparison.

Since with both types of connector there was only one set of recordings,
we could not draw any conclusion about the lower classification accuracy
obtained with the isolated montage. The vigilance level of the
participant alone could explain these performances.

Thanks to signals' correlations, however, we may infer that noise was
added to the system due to the presence of additional electrical
components in the adapter. Indeed, while the spectra were once again
strongly correlated, the averaged ERP achieved ``only'' a mean R score
of 0.88. Here external factors such as the metal state of the
participant or the quality of electrodes contacts could not have
influenced one amplifier rather than the other. Since temporal features
are more sensitive than spectral features to signal quality -- e.g.~one
``peak'' in the signal \emph{vs} oscillatory patterns over several
seconds --, it is instead more plausible that the difference with the
first experiment comes from the adapter.

Nonetheless, even though the ideal diode montage did not produce ideal
signals, those results still advocate for a close proximity between the
g.USBamp and the OpenBCI. No device behaved ``better'' than the other,
because no matter the lower correlation between averaged ERP, the
classification accuracy is in practice comparable between both
amplifiers. Each one probably endured different fluctuations since each
had a dedicated set of ideal diodes.

\section{CONCLUSION}\label{conclusion}

\noindent During this preliminary study, we compared the OpenBCI board
to the g.tec g.USBamp amplifier. We employed an original montage, based
on the simultaneous recording of the same set of electrodes. While as a
first approach we used a simple adapter with a direct connection between
the amplifiers and the electrodes, in a second experiment we attempted
to discard any possible interference that one amplifier could cause to
the other.

To do so, we built an adapter that embedded ``ideal'' diodes, components
that prevented electrical currents to flow ``backward''. This ensured
that we could test both devices in isolation. We did not try to compare
both adapters as the purpose was simply to gather more insights about
the possibility of simultaneous recordings -- this was a precaution to
detect a possible bias. For all applications and conditions AUROCC
scores were far beyond chancel level; the OpenBCI amplifier came close
to the g.USBamp in terms EEG features and effective performance.

That is not to say that the OpenBCI could replace an equipment such as
the g.USBamp, though. For instance, this open-hardware initiative does
not aim at medical applications, hence it should be employed in
sensitive contexts. It does not possess any certification; one reason
why so many cheap EEG device are wireless is not for practicality, but
to avoid any hazard due to power supply. Connecting somehow a body to
the power grid requires extra precautions and a certified isolation,
moreover when the impedance between the electrodes and the brain is
intentionally lowered.

There are also few issues with the current state of the OpenBCI project.
One concerns the sampling rate of the board. While 125Hz may be enough
for our use-cases -- no frequencies beyond 40Hz were used during the
work presented here -- it may not suffice others. The limitation of the
sampling rate is caused by the wireless protocol used for data
transmission. OpenBCI can deliver up to 250Hz signals to the computer,
but only on 8 channels instead of 16. Note that this may be optimized in
the future by updating the firmware or using alternate communications --
as far as the board itself is concerned, the documentation of the
ADS1299 chip ensuring analog-to-digital conversion claims a sampling
rate up to 16,000Hz.

With those limitations in mind, overall the results suggest that the
OpenBCI board -- or a similar solution based also on the Texas
Instrument ADS1299 chip -- could be an alternative to traditional EEG
amplifiers. Even though medical grade equipment possesses certification
and still outperforms the OpenBCI board in terms of classification, the
latter gives very close EEG readings. In practice, the obtained
classification accuracy may be suitable for reliable BCI in popular
settings, widening the realm of applications and increasing the number
of potential users.

\section{ACKNOWLEDGMENTS}\label{acknowledgments}

\noindent I thank Thibault Laine for his technical assistance during
this project.

\balance

\bibliographystyle{apalike}
{
\small
\bibliography{biblio.bib}}

\end{document}